\documentclass[epj]{svjour}
\usepackage{epstopdf}
\usepackage{physics}
\usepackage{tikz,lipsum,lmodern}
\usepackage[export]{adjustbox}
\usepackage{amsmath}
\usepackage{amssymb}

\usepackage[unicode=true,pdfusetitle,
 bookmarks=true,bookmarksnumbered=false,bookmarksopen=false,
 breaklinks=false,pdfborder={0 0 0},backref=false,colorlinks=true,citecolor=blue,urlcolor=violet 
]{hyperref}
\usepackage{xcolor}
\usepackage{graphicx}
\newcommand{\orc}[1]{\href{https://orcid.org/#1}{\includegraphics[width=10pt]{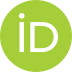}}}
\begin{document}

\title{Characterizing the generalized Einstein-Podolsky-Rosen state and extensions}
\titlerunning{Characterizing the generalized EPR state... }


\author{Rashi Adhikari\inst{1}\thanks{\email{rashi2305052@st.jmi.ac.in}}\orc{0009-0005-0873-883X} \and Mohd Shoaib Qureshi\inst{2}\thanks{\email{mohd2301184@st.jmi.ac.in}}\orc{0009-0003-4011-3173} \and Tabish Qureshi\inst{2}\thanks{\email{tqureshi@jmi.ac.in}}\orc{0000-0002-8452-1078}}
\institute{Department of Physics, Jamia Millia Islamia, New Delhi, India
\and
Centre for Theoretical Physics, Jamia Millia Islamia, New Delhi, India.
}

\abstract{
In their seminal paper, Einstein Podolsky and Rosen (EPR) had introduced a
momentum entangled state for two particles. That state, referred to
as the EPR state, has been widely used in studies on entangled particles
with continuous degrees of freedom. Later that state was generalized to a
form that allows varying degree of entanglement, known as the generalized
EPR state.  In a suitable limit it reduces to the EPR state. The generalized
EPR state is theoretically analyzed here and its entanglement quantified in
terms of a recently introduced generalized entanglement measure. This
state can also be applied to entangled photons produced from spontaneous
parametric down conversion (SPDC). The present analysis is then used
in quantifying the entanglement of photons produced from the SPDC process,
in terms of certain experimental parameters. A comparison is also made with
the Schmidt number, which is normally used as an entanglement measure in
such situations. A procedure for experimentally determining the entanglement
of SPDC photons has also been described. Furthermore, an additional state exhibiting non-Gaussian entanglement has been examined, and its entanglement has been quantified.}

\maketitle

\section{Introduction}

The phenomenon of entanglement \cite{schr} has attracted significant
attention in research due to its intriguing implications \cite{bell}.
It leads to effects that classical systems cannot replicate, thereby
establishing it as a fundamentally quantum phenomenon \cite{karimi}.
The practical applications of entanglement in fields such as quantum
cryptography \cite{ekert} and quantum teleportation \cite{teleport}
have further solidified its status as a distinct area of study \cite{rmp}.
The surprising and counter-intuitive aspects of entanglement were brought
to attention by Einstein Podolsky and Rosen (EPR) in their seminal paper
\cite{epr}.
\begin{figure}[h]
\centerline{\resizebox{8.5cm}{!}{\includegraphics{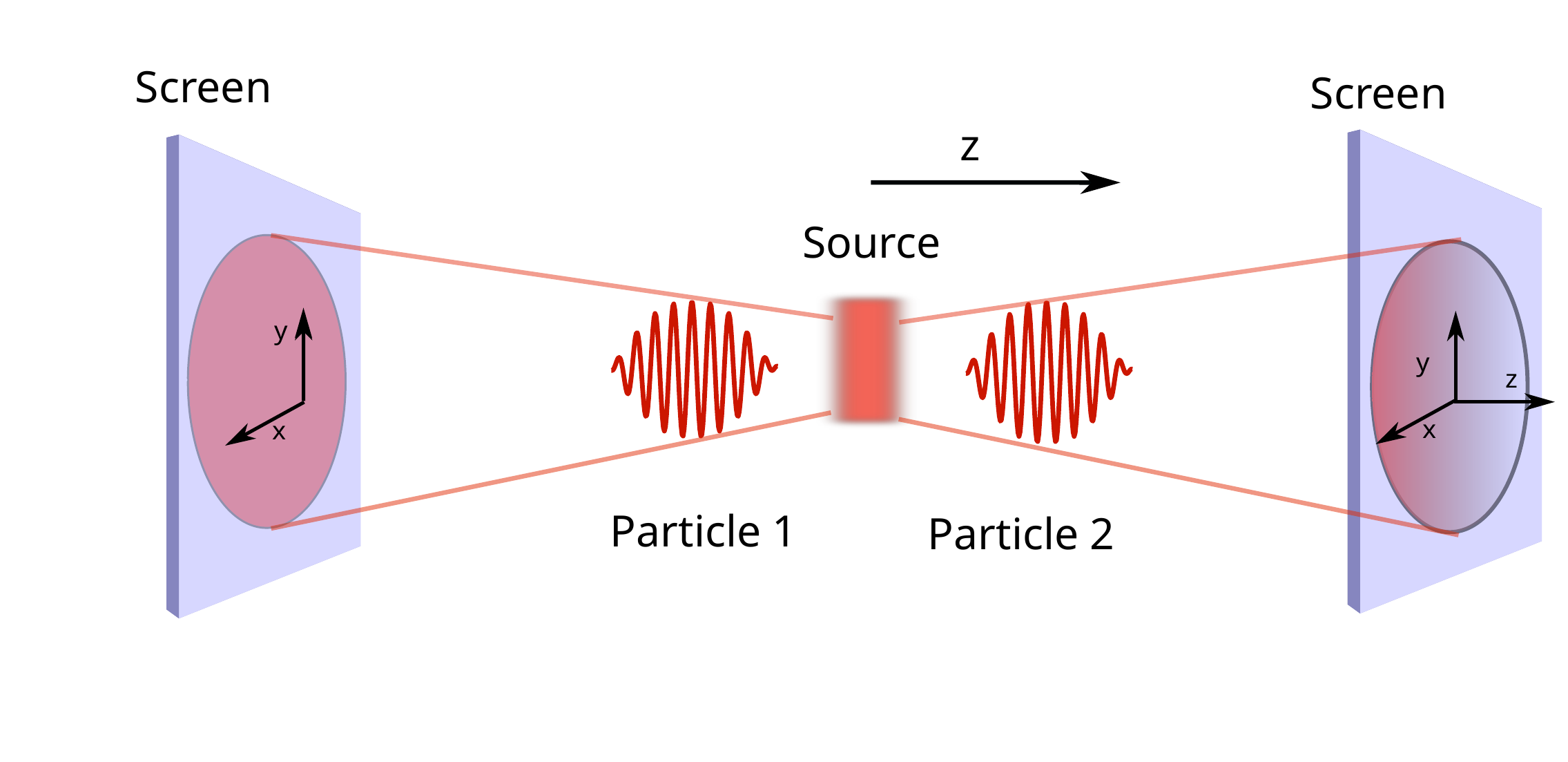}}}
\caption{Schematic diagram showing the generation of entangled particles.
The generated particles travel in opposite direction along z-axis.
We consider their entanglement in the transverse direction.}
\label{entparticles}
\end{figure}
Consider a source producing particles in pairs, which travel in opposite 
direction along the z-axis (see Fig. \ref{entparticles}). We are interested
in their entanglement in the transverse direction. The particles could,
for example, arise from the decay of a nucleus into two particles. They
could be electrons produced in pairs, in a specific process. They could
also be entangled photons. To simplify the situation
we may consider entanglement only along one direction, i.e., along x-axis.
When these two particles fall on the two screens, their detected positions
are found to be correlated. Needless to say, the momenta of the particles
are correlated too.
To analyze such entangled particles, Einstein Podolsky and Rosen introduced a
state, which has come to be known as the EPR state, and can
be written as \cite{epr}
\begin{equation}
\Psi_{EPR}(x_1,x_2) = \int_{-\infty}^\infty
e^{i px_1\over\hbar} e^{-{ipx_2\over\hbar}} dp, \label{epr}
\end{equation}
where $x_1$ and $x_2$ are the position variables of the two particles,
and $p$ is the momentum. The state can be thought of as two particles
moving with equal and opposite momenta, but forming a superposition of
such momenta.
This state captures the essential properties of entangled particles.
For a long time people believed that photon pairs generated from spontaneous
parametric down conversion (SPDC), are well described by the EPR state 
(\ref{epr}) \cite{klyshko}. Karl Popper had proposed an experiment to test
the Copenhagen interpretation of quantum mechanics \cite{popper}, which
relied on the implications of the EPR state. Popper's proposed experiment
became a focus of a long standing debate \cite{sudbery1,sudbery2,krips,collet,storey,redhead,popperreply}. The proposed experiment was realized
in 1999 \cite{popshih}, leading to a resurgence of controversy
\cite{nha,peres,hunter,sancho,shih,santo}. The controversy was resolved with the
recognition that in actual experimental conditions, momentum entangled
particles are not maximally entangled, and the EPR state is not suitable to
describe them \cite{tqajp,popreview}. While the EPR state does capture
the essential characteristics of momentum entangled
particles, it has some disadvantages like not being normalized, and not
describing varying degree of entanglement. The best state to describe
momentum-entangled particles is the {\em generalized EPR state}
\cite{tqajp,popreview,biphoton}
\begin{equation}
\Psi(x_1,x_2) = A\!\int_{-\infty}^\infty
e^{-{p^2\sigma^2\over \hbar^2}} e^{i px_1\over\hbar}e^{-{ipx_2\over\hbar}}
e^{-{(x_1+x_2)^2\over 4\Omega^2}} dp~, \label{gepr1}
\end{equation}
where $A$ is a normalization constant, and $\sigma,\Omega$ are certain
parameters. In the limit $\sigma\to 0,~~\Omega\to\infty$ the state (\ref{gepr1})
reduces to the EPR state (\ref{epr}).

After performing the integration over $p$, (\ref{gepr1}) reduces to
\begin{eqnarray}
\Psi(x_1,x_2) = \tfrac{1}{\sqrt{\pi\sigma\Omega}}
 e^{-(x_1-x_2)^2/4\sigma^2} e^{-(x_1+x_2)^2/4\Omega^2} .
\label{gepr}
\end{eqnarray}
It is straightforward to show that $\Omega$ and $\hbar/\sigma$ quantify
the position
and momentum spread of the particles in the x-direction because the
uncertainty in position and the wave-vector of the two particles,
along the x-axis, in the generalized EPR state, is given by
\begin{equation}
\Delta x_1 = \Delta x_2 = \sqrt{\Omega^2+\sigma^2},~
\Delta k_{1x} = \Delta k_{2x} = 
\tfrac{1}{4}\sqrt{\tfrac{1}{\sigma^2} + \tfrac{1}{\Omega^2}}~. \label{unc}
\end{equation}
It is evident that when $\sigma=\Omega$, the state (\ref{gepr}) becomes
disentangled; however, it is important to explore how the entanglement
of the state can be quantified in relation to the parameters $\sigma$ and
$\Omega$. Given that the generalized EPR state can represent the photon
pairs produced via the SPDC process \cite{eberly,walborn,howell},
this investigation will also facilitate the quantification of
entanglement in SPDC photons based on the experimental parameters. This
objective is the focus of the present investigation.

\section{Momentum entangled particles}

Quantifying the degree of entanglement of two particles is a much studied
subject \cite{plenio2016,bruss,danko}. For systems with finite degrees of
freedom, various entanglement measures have been defined. While all of
those cannot be listed here, we do mention some of them like
\emph{entanglement cost} \cite{EC1}, \emph{entanglement of formation}  
\cite{EF}, \emph{relative entropy of entanglement} \cite{REE},
\emph{logarithmic negativity} \cite{N}, \emph{entanglement coherence}
\cite{EC}. We particularly mention a measure
introduced by Hill and Wootters for two qubits, called \emph{concurrence}
\cite{con}, and its generalization to arbitrary Hilbert space dimensions,
called \emph{I-concurrence} \cite{rungta,pani}.
\begin{figure}[h]
\centerline{\resizebox{9.0cm}{!}{\includegraphics{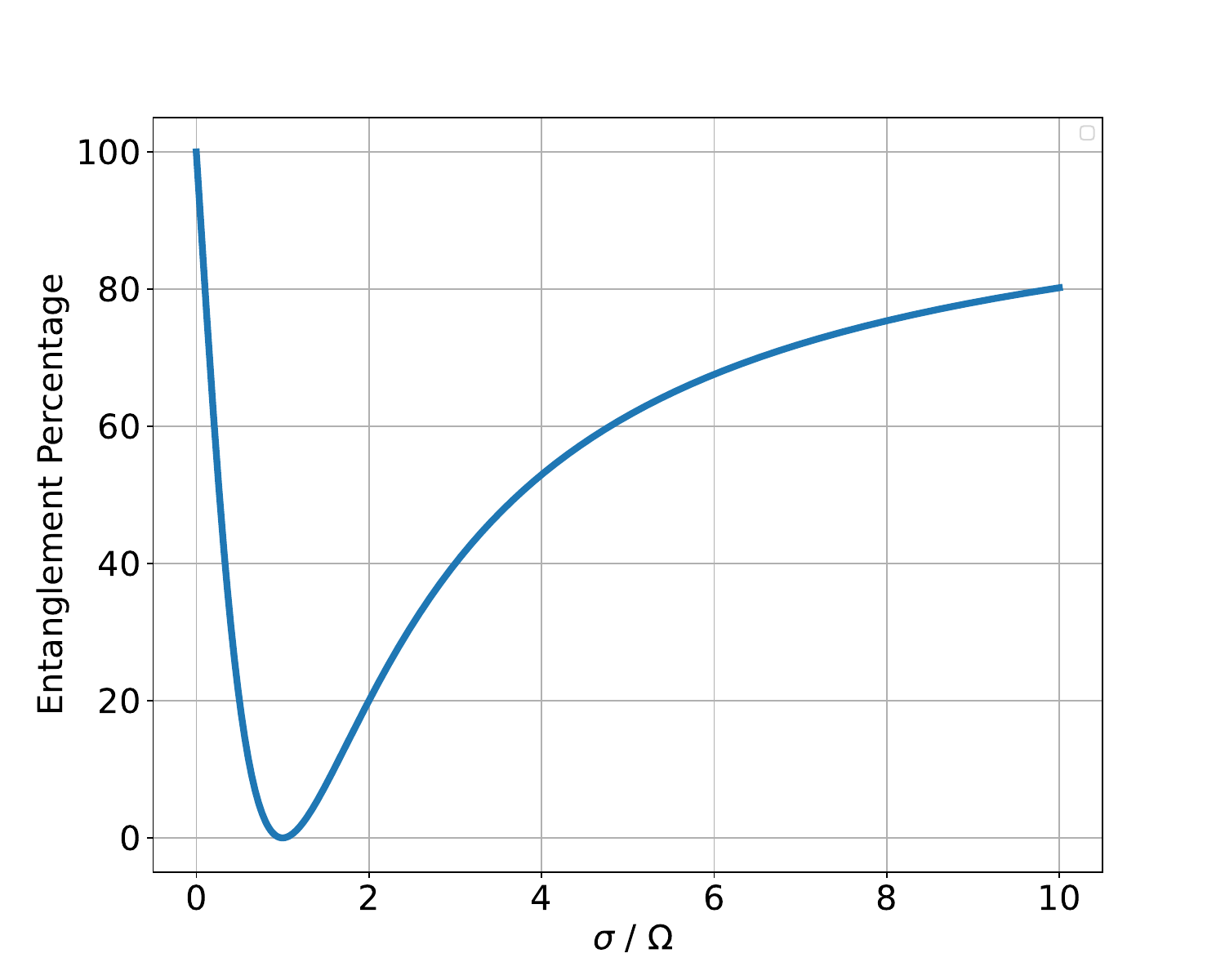}}}
\caption{The generalized entanglement measure plotted as percent entanglement
$\mathcal{E}^2\times 100/2$, against $\sigma/\Omega$.  }
\label{GEM}
\end{figure}

While the various entanglement measures mentioned above are very good in
their respective domains, they are not suited for quantifying entanglement
in the generalized EPR state, as it involves continuous variables. 
Quantifying entanglement for systems with continuous variables is different
ball game, and much effort has gone into it \cite{simon,agarwal,werner,kraus,duan,hillery,nha2}.
We find that a recently introduced generalized entanglement measure (GEM)
by Swain, Bhaskara and Panigrahi \cite{swain} is particularly suited for
analyzing the generalized EPR state.

Consider a $n-$partite entangled state
\begin{eqnarray}
|\psi\rangle = \int \psi(x_1,x_2,\dots,x_n)|x_1\rangle|x_2\rangle\dots|x_n\rangle dx_1 dx_2\dots dx_n ,\nonumber\\
\end{eqnarray}
with $\int\psi^*(x_1,x_2\dots,x_n)\psi(x_1,x_2\dots,x_n) d^nx = 1$.
Consider a bipartiton $\mathcal{M}$ of the combined Hilbert space, one
containing $m$ degrees of freedom, and the other containing $n-m$ degrees
of freedom. One may ask if the state is separable across this bipartition.
The GEM quantifies the degree of entanglement across this bipartition.
The GEM is defined for this $n-$partite entangled state, for the bipartition
$\mathcal{M}$, as \cite{swain}
\begin{eqnarray}
\mathcal{E}_{\mathcal M}^2 &=& 2\Big[1 - \int\Big|\int\psi(y_1',\dots,y_m',x_{m+1},\dots x_n)\nonumber\\
&&\psi^*(y_1,\dots,y_m,x_{m+1},\dots x_n)dx_{m+1}\dots dx_n\Big|^2
d^my d^my'\Big],\nonumber\\
\label{gemm}
\end{eqnarray}
The GEM is bounded by 0 and 2, in the spirit of I-concurrence. It is important
to point out that $\mathcal{E}^2_{\mathcal M}=0$ only implies that the state is
separable across this particular bipartition. There may still exist entanglement
within a partition. For our purpose, the bipartite case is relevant. Since
in a bipartite state there is only one bipartition possible,
the GEM gives the true entanglement measure for the state.
For the generalized EPR state (\ref{gepr}), the GEM can be evaluated, and
comes out to be
\begin{eqnarray}
\mathcal{E}^2 &=& 2 \frac{(\sigma-\Omega)^2}{\sigma^2+\Omega^2}
= 2 \frac{(\sigma/\Omega-1)^2}{\sigma^2/\Omega^2+1}
= 2 \frac{(1-\Omega/\sigma)^2}{1+\Omega^2/\sigma^2}.~~
\label{gem}
\end{eqnarray}
Firstly, for $\sigma=\Omega$ the state is disentangled, and
$\mathcal{E}^2=0$. For a nonzero $\Omega$ we have
\begin{eqnarray}
\lim_{\sigma\to 0}\mathcal{E}^2 = \lim_{\sigma\to 0}2\frac{(\sigma/\Omega-1)^2}{\sigma^2/\Omega^2+1} = 2,
\end{eqnarray}
which implies that the state is maximally entangled. In this situation, if
particle 1 is detected at $x_1=x_0$, particle 2 will also be found at
$x_1=x_0$. For a nonzero $\sigma$ we have
\begin{eqnarray}
\lim_{\Omega\to 0}\mathcal{E}^2 = \lim_{\Omega\to 0}2\frac{(1-\Omega/\sigma)^2}{1+\Omega^2/\sigma^2} = 2,
\end{eqnarray}
which also implies that the state is maximally entangled. In this situation, if
particle 1 is detected at $x_1=x_0$, particle 2 will be found at $x_1=-x_0$,
which means that the two particles are oppositely correlated in position. 
In practice neither $\sigma$ nor $\Omega$ can actually go to zero. 
The two limits of strong entanglement will be more like
$\sigma\ll\Omega$ and $\sigma\gg\Omega$.
The nice thing about having a well defined upper bound for GEM is that for
any given values of $\sigma$ and $\Omega$, one can estimate the percentage
of entanglement present in the quantum state. {A word of caution is in order here. While a normalizable entanglement measure does allow us to talk of the degree of entanglement in terms of percentage, it is closely tied to the particular entanglement measure one is using. A different normalized entanglement measure may yield a different degree of entanglement for the same state.}
\begin{figure}[h]
\centerline{\resizebox{9.0cm}{!}{\includegraphics{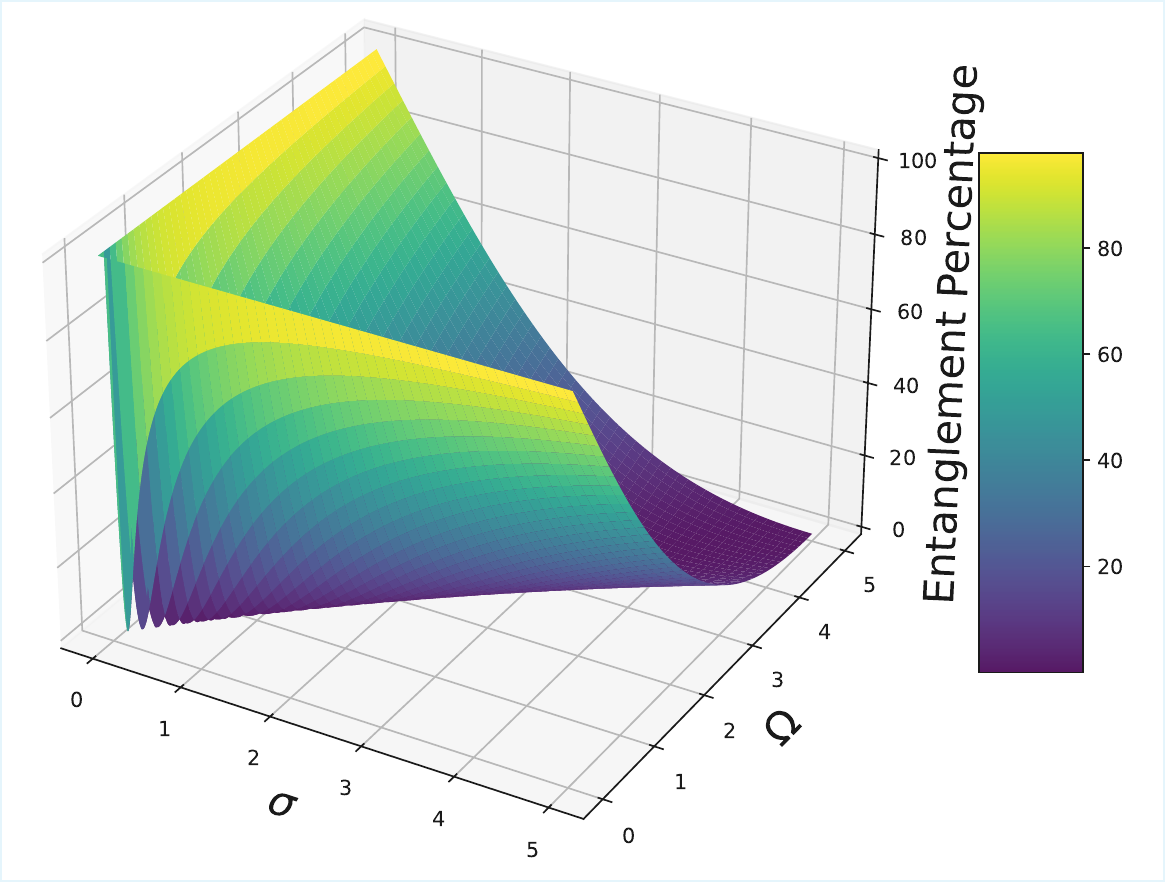}}}
\caption{The generalized entanglement measure plotted as percent entanglement
$\mathcal{E}^2\times 100/2$, against $\sigma$ and $\Omega$.}
\label{GEM_3D}
\end{figure}

Fig. \ref{GEM} shows the GEM plotted against $\sigma/\Omega$. The plot
with an obvious dip at $\sigma/\Omega=1$, shows how the two parameters can be
tuned to achieve a desired degree of entanglement. For example, for the case 
$\sigma >\Omega$, tuning the parameters to $\sigma/\Omega=10$ can give
one 80\% entanglement. Behavior of the GEM with the two
parameters varying independently, is depicted in Fig. \ref{GEM_3D}.
It represents the complete characterization of entanglement in the
generalized EPR state.

\section{SPDC photons}

\begin{figure}[h]
\centerline{\resizebox{8.5cm}{!}{\includegraphics{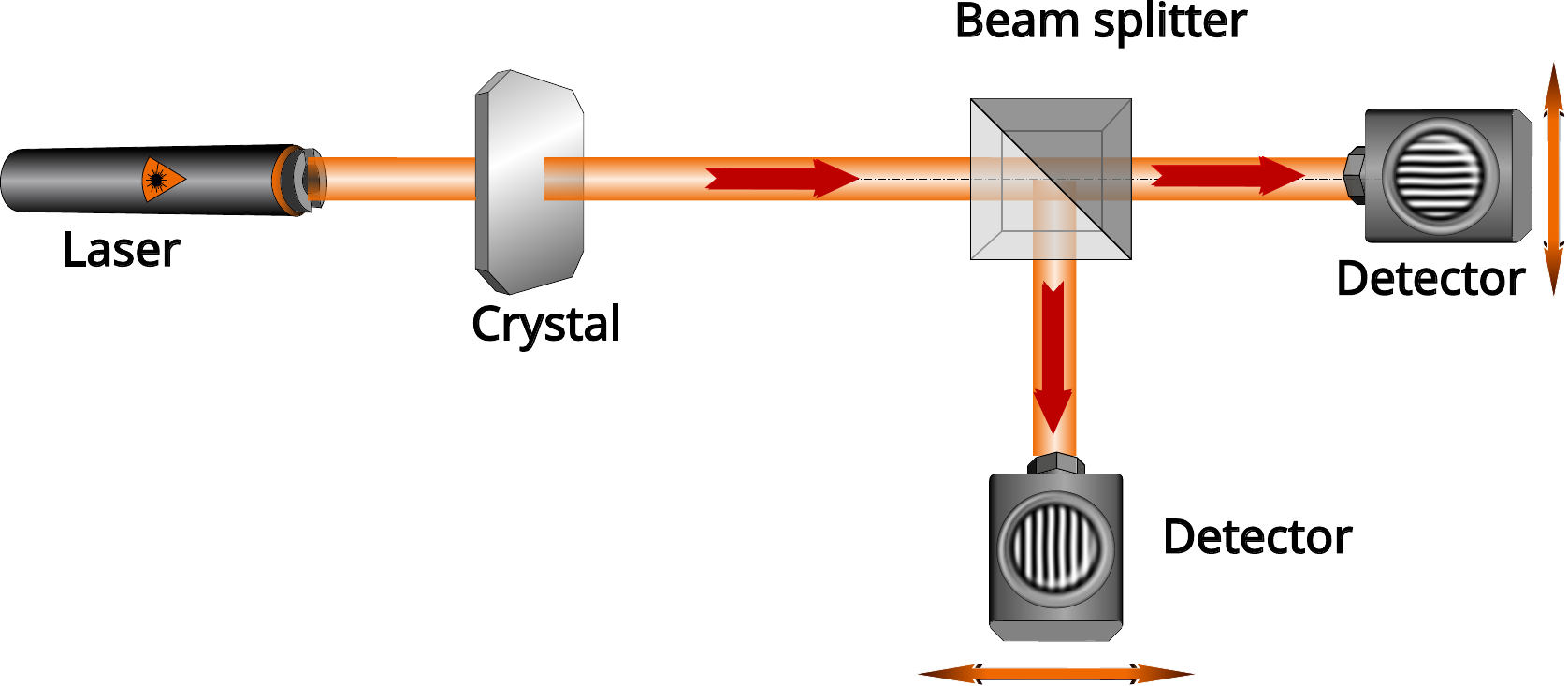}}}
\caption{Schematic diagram showing the generation of SPDC photons. A photon
from a laser falls on a nonlinear crystal, resulting in the emission of
two photons which are entangled. The two photons are typically separated
via a polarization beam splitter.}
\label{spdc}
\end{figure}

The process of spontaneous parametric down-conversion (SPDC) is a
nonlinear optical phenomenon that takes place in birefringent
crystals, where high-energy 'pump' photons are transformed into pairs
of lower-energy 'signal' and 'idler' photons. Specifically, the pump
field engages in coherent interaction with the electromagnetic quantum
vacuum through a nonlinear medium, resulting in the annihilation of
a pump photon and the simultaneous creation of two daughter photons
(signal and idler), a process that occurs repeatedly (see Fig. \ref{spdc}).
As this is a
parametric process—characterized by the initial and final states
of the crystal remaining unchanged—both the total energy and total
momentum of the field must be conserved. Consequently, the energies and
momenta of the daughter photons exhibit a high degree of correlation,
leading to a highly entangled quantum state.

Photons generated from SPDC process are routinely used as entangled particles.
The theory behind the SPDC process is now well established
\cite{eberly,walborn,howell}. The important experimental parameters that
effect the entanglement of the two photons are $\sigma_p$, the pump beam
radius, $\lambda_p$, the wavelength of the pump beam, and $L$ the thickness
of the nonlinear crystal along the propagation direction of the beam.
In degenerate, collinear SPDC, the momentum-space wave function of the biphoton
can be written as \cite{howell}
\begin{eqnarray}
\Phi(\vec{k_1},\vec{k_2}) = \mathcal{N}\text{sinc}\left(\frac{L\lambda_p}{8\pi}
|\vec{q}_1-\vec{q}_2|^2\right) e^{-\sigma_p^2|\vec{q}_1+\vec{q}_2|^2} ,
\label{spdcp}
\end{eqnarray}
where $\text{sinc}(\theta)\equiv \sin(\theta)/\theta$, $\vec{q}_1$, and $\vec{q}_2$ are the projections of the 
signal wave vector $\vec{k}_1$, and the idler wave vector $\vec{k}_2$, onto
the plane transverse to the optic axis, respectively. Here a Gaussian
profile of the pump beam is assumed. The sinc function can be approximated
as a Gaussian, under certain conditions, and (\ref{spdcp}) can be
inverse-Fourier transformed to give the following position space wave function,
along the x-direction, for the biphoton:
\begin{eqnarray}
\psi(x_1,x_2) = \mathcal{A}e^{-\frac{3\pi}{2L\lambda_p}
(x_1-x_2)^2} e^{-(x_1+x_2)^2/16\sigma_p^2 } ,
\label{spdcx}
\end{eqnarray}
\begin{figure}[h]
\centerline{\resizebox{8.5cm}{!}{\includegraphics{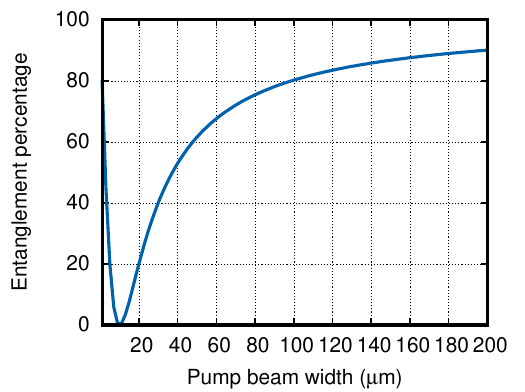}}}
\caption{Variation of entanglement of two SPDC photons for
$\sqrt{\frac{L\lambda_p}{6\pi}} \approx 0.01~\text{mm}$, as a functions of
pump beam width $2\sigma_p$.}
\label{spdcph}
\end{figure}
$\mathcal{A}$ being the normalization constant.
This position-space wave function of the biphoton is of the exact same form 
as the generalized EPR state (\ref{gepr}), with the two parameters of
the generalized EPR state identified with
\begin{eqnarray}
 \sigma = \sqrt{\frac{L\lambda_p}{6\pi}}~~~\text{and}~~~ \Omega = 2\sigma_p .
\label{sigmaomega}
\end{eqnarray}
So physically, parameter $\Omega$ is just the width of the pump beam.
Using the above, the generalized entanglement measure of the biphoton can be
written as
\begin{eqnarray}
\mathcal{E}^2 &=& 2 \frac{(\sqrt{L\lambda_p/24\pi\sigma_p^2}-1)^2}{L\lambda_p/24\pi\sigma_p^2+1}.
\label{gemphoton}
\end{eqnarray}
Relation (\ref{gemphoton}) can be of much practical use. Knowing the
parameters of the SPDC setup, one can estimate the degree of entanglement
of the photons produced. As an example, we consider a recent experiment
in which a 405 nm
horizontally polarized pump beam was spatially filtered and directed through
a Type-II PPKTP (Periodically Poled Potassium Titanyl Phosphate) crystal of 
thickness $L = 10$ mm \cite{kiran}. The pump beam width was set to $\approx$
350 $\mu$m. Using (\ref{gemphoton}) with these values, the GEM comes out
to be $\mathcal{E}^2 = 1.832$, which indicates that the photons are 
91.6\% entangled. An SPDC experiment using an unconventionally thick crystal
surprisingly demonstrated generation of entangled photons \cite{Septriani}.
In this experiment, a $\beta$-barium borate crystal of thickness $L=15.76$ mm
was used, with a
405 nm laser having a Gaussian beam thickness of 180 $\mu$m. These values
yield a GEM of $\mathcal{E}^2 = 1.796$, corresponding to 89.8\% entanglement.
One may also like to study how the degree of entanglement
varies as a function of the pump beam width $\sigma_p$. In many SPDC
experiments, typically
$\sigma = \sqrt{\frac{L\lambda_p}{6\pi}} \approx 0.01~\text{mm}$ \cite{howell}.
Fig. \ref{spdcph} shows how the entanglement of the two photons vary as
a function of pump beam width $\sigma_p$.

It is pertinent to mention here an entanglement measure frequently used for
entangled photons, namely, the Schmidt number \cite{schmidtn}. This measure
is based on Schmidt decomposition of entangled states \cite{schmidt}.
The Schmidt number $K$ provides the ``average" number of Schmidt modes
involved in an entangled state. For the entangled state of SPDC photons,
under consideration here, the Schmidt number has a closed form \cite{schmidtn}
\begin{eqnarray}
{K} &=& \frac{1}{4}\left(\frac{\sigma}{\Omega} + \frac{\Omega}{\sigma}\right)^2 .
\label{schmidtgepr}
\end{eqnarray}
While for a disentangled state, $K=1$, the upper limit of $K$ is
determined by the volume of available phase space within the system,
constrained by pertinent physical laws such as the conservation of energy
and momentum. In this respect, the GEM provides a more convenient way to
gauge how far a state is from maximal entanglement.

\section{Measurement of entanglement}

The question we finally address is, given pairs of entangled photons, how
does one experimentally measure their entanglement. In the present context,
it amounts to measuring the GEM. The GEM for entangled photons
(\ref{gemphoton}) can be written as
\begin{eqnarray}
\mathcal{E}^2 &=& 2\left(1 - \sqrt{\frac{2}{\sigma^2+\Omega^2}} \sqrt{\frac{2\sigma^2\Omega^2}{\sigma^2+\Omega^2}}\right)\nonumber\\
&=& 2\left(1 - \frac{\sigma_{(1|2)}}{\sigma_1}\right) ,
\end{eqnarray}
where $\sigma_1 = \sqrt{\frac{\sigma^2+\Omega^2}{2}}$ is the marginal and
$\sigma_{(1|2)} = \sqrt{\frac{2\sigma^2\Omega^2}{\sigma^2+\Omega^2}}$ the
conditional of the double Gaussian distribution $P(x_1,x_2)$ corresponding
to (\ref{spdcx}). 
It has recently been shown that the correlation width of a biphoton state
$\psi(x_1,x_2)$ can be determined by inversion interferometry \cite{jha}.
The interference cross term, represented by the one-photon cross-spectral
density function, is
\begin{eqnarray}
W(x_1,x_1') = \int \psi(x_1,x_2)\psi(x_1',x_2) dx_2.
\end{eqnarray}
It has been shown that under certain conditions $W(x_1,-x_1) \propto
P(x_1|x_2=0)$ \cite{jha}. 
Consequently, the conditional distribution $P(x_1,x_2=0)$ has the same
width as the
cross-spectral density. The width $\sigma_{(1|2)}$ of $P(x_1,x_2=0)$
can be obtained by measuring the width $f$ of $W(x_1,-x_1)$, which is
easier since no coincidence measurements are required \cite{jha}.
Since the marginal distribution here is the downconverted beam, $\sigma_1$
is just the width of the beam. Thus, the GEM is given by
\begin{equation}
\mathcal{E}^2 = 2\left(1 - \frac{f}{\sigma_1}\right) ,
\end{equation}
where $f$ and $\sigma_1$ are obtained experimentally as described above.
The experimental determination of $f$ and $\sigma_1$ has already been
demonstrated by Bhat et al. \cite{bhat}.

{\section{Going beyond Gaussian entanglement}}

{Historically, the majority of focus on continuous variable systems
in quantum information processing has been directed towards Gaussian
states. Nevertheless, the presence of non-Gaussianity is a crucial
prerequisite for achieving universal quantum computation and entanglement
distillation, and it can enhance the effectiveness of various other
quantum information tasks \cite{plenio2002,illuminati,cloning,gomes}.
Gaussian entanglement can be detected by using the continuous variable
equivalent of Peres-Horodecki separability criterion \cite{simon}
\begin{equation}
\big\langle \delta \hat{x}_1\delta \hat{x}_2\big\rangle
\big\langle \delta \hat{p}_1\delta \hat{p}_2\big\rangle
-\big\langle \delta \hat{x}_1\delta \hat{p}_2\big\rangle
\big\langle \delta \hat{p}_1\delta \hat{x}_2\big\rangle \ge 0,
\label{PH}
\end{equation}
where $\delta \hat{x} = \hat{x}-\langle\hat{x}\rangle$,
$\delta \hat{p} = \hat{p}-\langle\hat{p}\rangle$.
If the above condition holds, second order moments cannot detect
entanglement, and the state appears to be separable. For Gaussian
states, this condition is sufficient to detect entanglement. In order
to convince the reader that the generalized EPR state (\ref{gepr})
falls in this category, we evaluate the condition (\ref{PH}) for it:
\begin{eqnarray}
\big\langle \delta \hat{x}_1\delta \hat{x}_2\big\rangle
\big\langle \delta \hat{p}_1\delta \hat{p}_2\big\rangle
-\big\langle \delta \hat{x}_1\delta \hat{p}_2\big\rangle
\big\langle \delta \hat{p}_1\delta \hat{x}_2\big\rangle =
-\tfrac{\hbar^2}{16} (\tfrac{\sigma}{\Omega}-\tfrac{\Omega}{\sigma})^2. \nonumber\\
\end{eqnarray}
Clearly, the generalized EPR state violates the separablity criteria unless
$\sigma=\Omega$.}
\begin{figure}[h]
\centerline{\resizebox{9.0cm}{!}{\includegraphics{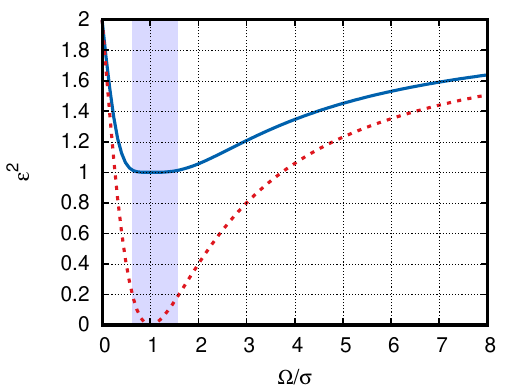}}}
\caption{{The generalized entanglement measure for the non-Gaussian state
(\ref{ngauss}) (solid line) and for the generalized EPR state (\ref{gepr}) (dashed line), plotted
against $\Omega/\sigma$. The shaded region represents the parameter range
$0.63 < \Omega/\sigma < 1.58$ where the entanglement of the state (\ref{ngauss}) is non-Gaussian. }}
\label{GEMNG}
\end{figure}

{Let us now consider the following state which can be shown to possess
non-Gaussian entanglement \cite{gomes}
\begin{eqnarray}
\Psi_{NG}(x_1,x_2) = \tfrac{x_1+x_2}{\sqrt{\pi\sigma\Omega^3}}
 e^{-\frac{(x_1-x_2)^2}{4\sigma^2}} e^{-\frac{(x_1+x_2)^2}{4\Omega^2}} .
\label{ngauss}
\end{eqnarray}
It merits consideration here because it is an example of an experimentally
accessible non-Gaussian state. It can be verified that (\ref{ngauss})
satisfies the criterion (\ref{PH}) provided $0.57 < \Omega/\sigma < 1.73$,
even though the state is clearly entangled. This means that no second order
criteria can detect the entanglement of this state, in the given parameter
range. However, it has been shown that there is a \emph{higher-order
separability criterion} that this state violates for the parameter range
$0.63 < \Omega/\sigma < 1.58$ \cite{gomes}. So in this range no second-order
criteria can detect the entanglement of the state (\ref{ngauss}), while a
higher order criterion does. This is a clear demonstration of non-Gaussian
entanglement.}

\begin{figure}[h]
\centerline{\resizebox{9.0cm}{!}{\includegraphics{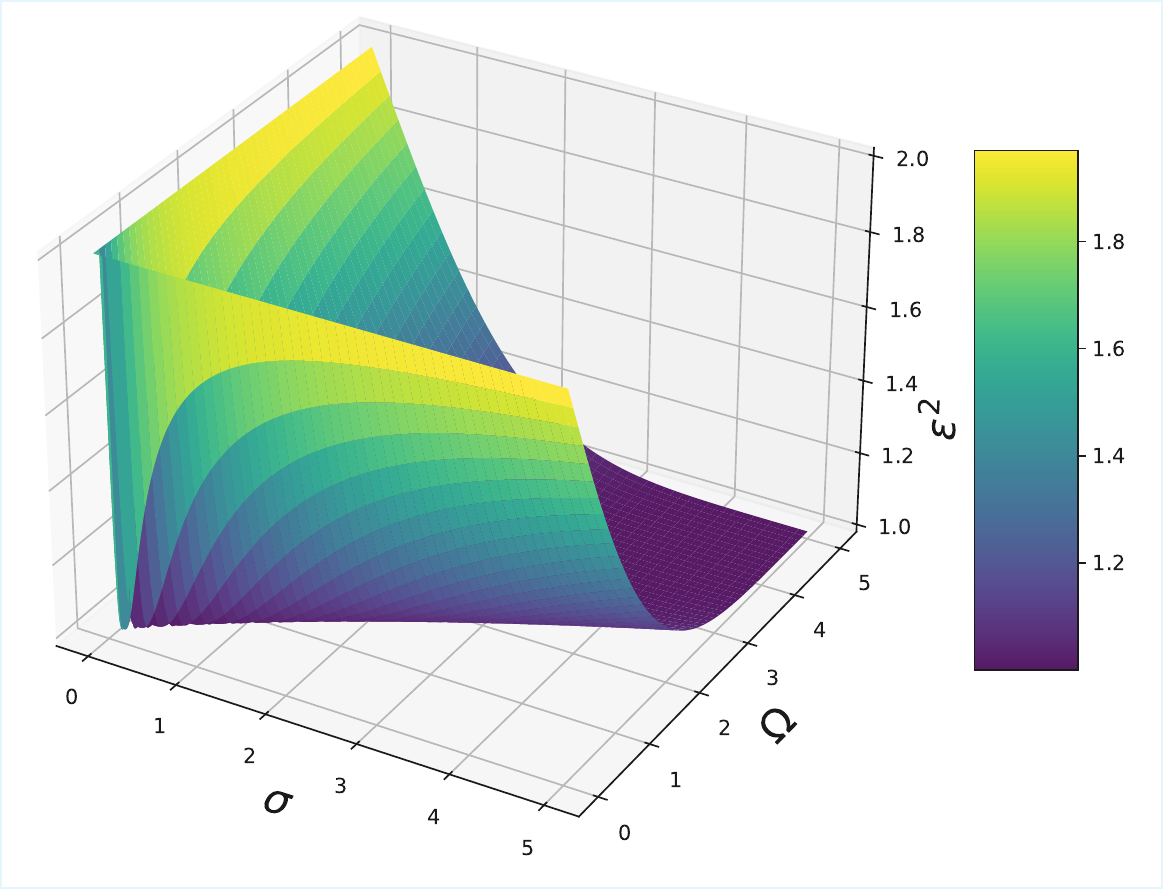}}}
\caption{{The generalized entanglement measure for the non-Gaussian state (\ref{ngauss}) plotted against $\sigma$ and $\Omega$.}}
\label{GEM_NG_3D}
\end{figure}

{For the non-Gaussian state (\ref{ngauss}), the GEM, as given by
(\ref{gemm}), can be evaluated, and after considerable algebra, comes out to be
\begin{eqnarray}
\mathcal{E}^2 &=& 2-\frac{\sigma\Omega(3\Omega^4+2\Omega^2\sigma^2+3\sigma^4)}
{(\sigma^2+\Omega^2)^3}.
\label{gemng}
\end{eqnarray}
For $\Omega=\sigma$ the state (\ref{ngauss}) is still entangled, and $\mathcal{E}^2 =1$. The state is maximally entangled in two limits, $\sigma\to 0$ and $\Omega\to 0$.
The GEM for the state (\ref{ngauss}) has been plotted in Fig. \ref{GEMNG} and Fig. \ref{GEM_NG_3D}. The first thing to notice is that the state does not become disentangled for any parameter values. Secondly, for any  value of $\Omega/\sigma$ the entanglement in the non-Gaussian state is higher than that in the generalized EPR state.
Interestingly, the GEM is the lowest in the parameter range for which the
entanglement is strictly non-Gaussian.}

\section{Conclusion}

In conclusion, we quantified the entanglement of the generalized EPR state
using a recently introduced measure of entanglement suitable for continuous
variables, namely, the generalized entanglement measure (GEM). The entanglement
measure can be written in a closed form in terms of the two parameters of
the generalized EPR state. Since the generalized EPR state can also describe
entangled photons produced from the SPDC process, we quantified the
entanglement of SPDC photons in terms of the relevant experimental parameters.
Unlike the commonly used measure of entangled photons, Schmidt number, the
GEM is able to estimate the \emph{percentage of entanglement} present in SPDC
photon state. We believe this quantification would be very useful in 
experiments with entangled photons. We also described how GEM can be measured
experimentally for SPDC photons.
{We also derived a closed expression for the entanglement measure of a non-Gaussian state. This may prove useful in understanding the relationship of such measures with the non-Gaussianity of entanglement.}


\vspace{5mm}
\emph{No additional data was generated in this work.}

\section*{Author contribution}

Tabish Qureshi conceptualized the research problem and wrote the major
part of the manuscript. Rashi Adhikari and Mohd Shoaib carried out the
calculations and generated the plots.

\end{document}